# Dynamic short-range correlation in photoinduced disorder phase transitions

Wen-Hao Liu[1,2], Jun-Wei Luo[1,2]*, Shu-Shen Li[1,2], and Lin-Wang Wang[3]*

[1]*State Key Laboratory of Superlattices and Microstructures, Institute of Semiconductors, Chinese Academy of Sciences, Beijing 100083, China*

[2]*Center of Materials Science and Optoelectronics Engineering, University of Chinese Academy of Sciences, Beijing 100049, China*

[3]*Materials Science Division, Lawrence Berkeley National Laboratory, Berkeley, California 94720, United States*

*Email: jwluo@semi.ac.cn; lwwang@lbl.gov

**Abstract:**

Ultrafast photoexcitation can induce a nonequilibrium dynamic with electron-lattice interaction, offering an effective way to study photoinduced phase transitions (PIPTs) in solids. The issue that atomic displacements after photoexcitation belong to coherent change or disordered process, has become a controversy in the PIPT community. Using real-time time-dependent density functional theory (rt-TDDFT) simulations, we obtain both the coherent and the disordered PIPTs (dimer dissociation) in $IrTe_2$ with the different electronic occupations. More importantly, we found that in the disordered phase transition, there exists a local correlation between different dimers regarding their dissociation status. One can define vertical groups across the layers. The dimers in the same group will dissociate in a correlated fashion: they either all dissociate, or all not dissociate. On the other hand, the dimers in neighboring groups will have an anti-correlation: if the dimers in one group dissociate, the dimers in the neighboring group tend not to be dissociated, and vice versus.

**Introduction**

Photoinduced phase transitions (PIPTs) originate from a complex interplay of instabilities in the charge, spin, orbital, and lattice degrees of freedom, and thus the understanding of interactions among different degrees of freedom is an essential key for unraveling and controlling material properties [1-10]. Recently, the issue of coherent and correlated change [2,11-17], versus disordered dynamics after photoexcitation [18-20], becomes a critical debate in the understanding of the ultrafast phase transition process. The coherent phonon following photoexcitation is always treated as the crucial factor to induce atomic motion [10,15,16]. For example, in $Me_4P[Pt(dmit)_2]_2$, it is directly observed that the coherent displacement of a collective phonon mode leads to the evolution of Pt-Pt dimers with PIPT [14]. However, in a recent experiment related to the $VO_2$ phase transition, it is shown that the movement of different V-V dimers becomes independent without correlation immediately after photoexcitation, long before the material heats up due to the transfer of energy from electron degree of freedom to the lattice vibration [18,19].

$IrTe_2$ as a special quasi-2D layered transition-metal ditellurides includes a high-temperature (HT) phase with high symmetry and a low-temperature (LT) phase with $q_{1/5}$ = (1/5, 0, 1/5) dimerized modulation, as shown in Fig. 1(a) [21-23]. The dissociation of Ir-Ir dimers is a direct descriptor for LT-to-HT phase transition for $IrTe_2$ [24-26]. Detailed and deeper theoretical study of the dimerization evolution can shed some light on this critical problem for general ultrafast phase transitions [26-28]. Previously, we have theoretically studied the microscopic mechanism and atomic forces of Ir-Ir dimer dissociation [26] and the effect of hot carrier cooling on the PIPT time scales [27]. However, the issue of coherence versus disorder phase transition has not been studied.

In this work, through real-time time-dependent density functional theory (rt-TDDFT) simulations, we study the dynamical process of excited electrons to the conduction band, in particular the coherent or disorder movement problem. When 3% of valence electrons are manually placed to the bottom of the conduction band and the Ir-Ir dimerized antibonding states are 100% occupied by the excited electrons, the Ir-Ir dimers will be coherently dissociated in 300 fs (Fig. 1(b)). On the other hand, when the laser pulse is used to excite the same 3% electrons to the conduction band, the dimerized antibonding states are only 40% occupied, and the other electrons are placed in higher-energy conduction band states. Due to the smaller driving force, the thermal phonons become more important, and the coherence of the atomic movement will lose quickly, which leads to disordered vibration in the dimer as shown in Fig. 1(c). This is similar to the experiment in $VO_2$ [18], and the dimers will only be partially dissociated in a much longer timescale of about 2.0 ps (only one dimer dissociated in the dimer group in Fig. 1(a)). Interestingly, in the disordered dynamics, we found a correlation for the dissociation process between the dimers. Regarding whether a dimer dissociates or does not dissociate, one can define vertical cross-layer dimer groups as shown in Fig.1. The dimers in the same group have a positive correlation for their dissociation/non-dissociation status, while dimers in neighboring groups have an anti-correlation. More specifically, this means the dimers in the same group will either all be dissociated, or none be dissociated. On the other hand, the dimers in groups I and II, if dimers in I are dissociated, then the dimers in II will not be dissociated, vice versus. Thus, this is a local short-range order for the dynamic process in the disordered PIPTs.

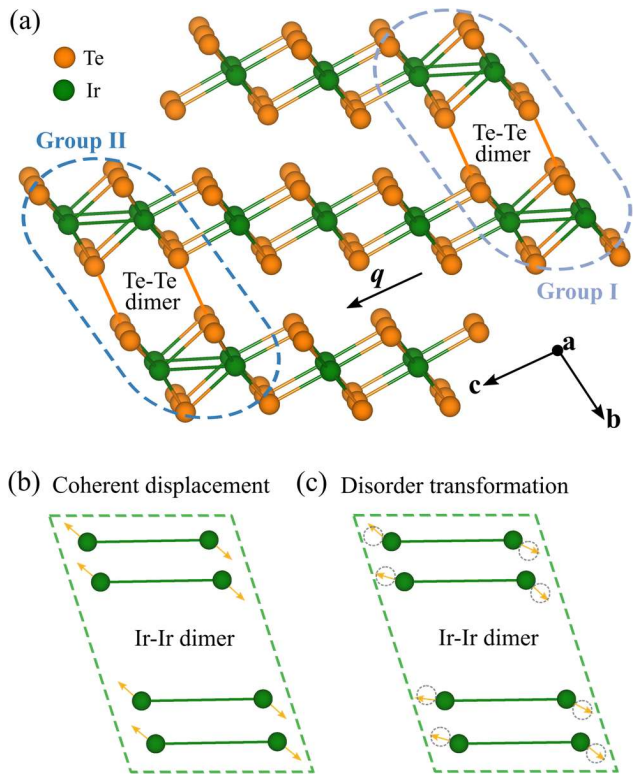

**Fig. 1. Schematic diagram of LT phase in $IrTe_2$.** (a) LT phase with Ir-Ir dimerized $q = (1/5, 0, 1/5)$ lattice modulation. The 120-atom supercell used in simulations includes eight pairs of Ir-Ir dimers which are classified into two groups (I and II) with the same symmetry. (b) The coherent displacement of Ir-Ir dimers. (c) The disorder transformation of Ir-Ir dimers.

**Method**

We carry out the rt-TDDFT simulations [29] based on the norm-conserving pseudopotentials (NCPP) [30] and Perdew-Burke-Ernzerhof (PBE) functional within density functional theory (DFT) framework with a plane wave nonlocal pseudopotential Hamiltonian, which is implemented in the code PWmat [31]. The wave functions were expanded on a plane-wave basis with an energy cutoff of 45 Ryd. In the rt-TDDFT simulation, we use a 128-atom supercell for $IrTe_2$ and the $\Gamma$ point is used to sample the Brillouin zone. The time step is set to 0.1 fs in our simulations. In this work, we apply two methods to achieve electronic excitation. We first manually place 3% of valence electrons to the bottom of the conduction band in Fig. 2(a). Then, to simulate the photoexcitation, we apply an external electric field to represent a laser pulse with a Gaussian shape during rt-TDDFT simulation: $E(t) = E_0 \cos(\omega t) \exp[-(t-t_0)^2/(2\sigma^2)]$, where $E_0 = 0.2$ V/Å, $t_0 = 60$ fs, $2\sigma = 25$ fs is the pulse width, and $\omega = 3.1$ eV is the photon energy [25]. We obtain the 3% photoexcited electrons, as shown in Fig. 2(d).

# Results

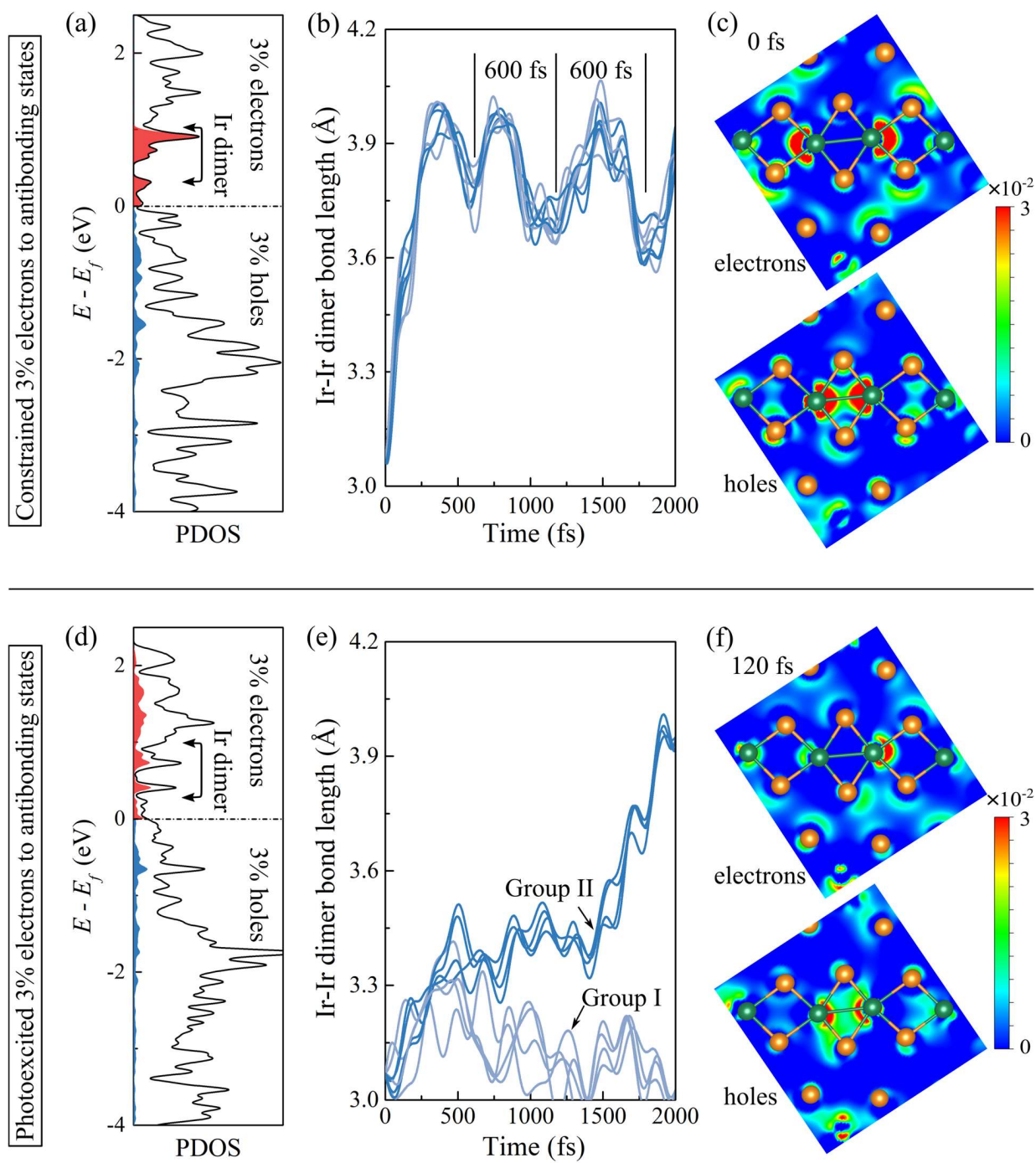

**Fig. 2. Dynamics upon constrained and photoexcited 3% electrons to antibonding states.** (a-c) for the case of constrained 3% electrons to antibonding states. (d-f) for the case of photoexcited 3% electrons to antibonding states. (a), (e) Partial density of states (PDOS). Red (blue) shaded areas represent the excited electrons (holes). (b), (e) Time dynamics of eight-pair Ir-Ir dimerized bond length. The shallow (dark) blue lines represent Ir-Ir

dimers in group I (II) as labeled in Fig. 1. (c), (f) Real-space distributions of the excited electrons and holes on the $(11\bar{1})$ plane.

First, to obtain the fast phase transition, we manually place 3% of valence electrons to the bottom of the conduction band where the Ir-Ir dimerized antibonding states are 100% occupied by the excited electrons, as shown in Fig. 2(a). Afterward, we proceed with a TDDFT dynamic simulation with a 3% excitation and the system remains at a lower temperature (~ 200 K) before the material heats up due to energy transfer from the electron degree of freedom. The bond length of all Ir-Ir dimers increases quickly from an initial value of 3.1 Å to 3.9 Å within 300 fs—half a vibrational period of phonons (~ 1.6 THz) in this system, exhibiting the breaking of the Ir-Ir dimers (Fig. 2(b)). The phenomenon is similar to the non-thermal melting of bulk bismuth at high excitation densities which can be driven by coherent phonon in 190 fs—half a period of the unperturbed $A_{1g}$ lattice mode [11]. This kinetic process is in a collective coherent mode without large random fluctuation (Fig. 2(b)). The fast dissociation of the Ir-Ir dimers is attributed to the occupations of its antibonding states by excited electrons and the bonding states by excited holes, which generates atomic driving forces to stretch the Ir-Ir dimer bonds to lower the free energy of the excited system [26]. In the real-space plots of the photoexcited electrons and holes shown in Fig. 2(c), we can see that the holes are located at the center of the Ir-Ir dimer bond, while the excited electrons are located at the outside ends of the Ir-Ir dimer bond. This is consistent with our previous analysis [26], which explains the coherent driving forces that dissociate the Ir-Ir dimers.

Next, we apply a 400-nm 120-fs laser pulse, with the laser strength adjusted to excite 3% valence electrons to conduction bands as shown in Fig. 2(d). Here, only 40% Ir-Ir dimerized antibonding states are occupied by the excited electrons, and other photoexcited electrons occupy the higher energy states. The lattice temperature remains around 200 K during the simulation. The dissociation of Ir-Ir dimers doesn't happen in a short time due to the lower occupations of Ir-Ir dimerized antibonding states. However, eventually, the dimers will have partial dissociation at about 2.0 ps as shown in Fig. 2(e). Compared with Fig. 2(c), the holes still concentrate at the center of the Ir-Ir bond, and the electrons still concentrate at the two outer ends of the Ir-Ir bond, but the patterns are less apparent (Fig. 2(f)). As a result, the coherent kinetic driving forces for dimer dissociation are much smaller, which leads to a much longer dissociation time (and only with partial dissociation). The timescale of Ir-Ir dimer dissociation in photoinduced simulation is 3 times slower than constrained electrons on the bottom of the conduction band case.

**Discussion**

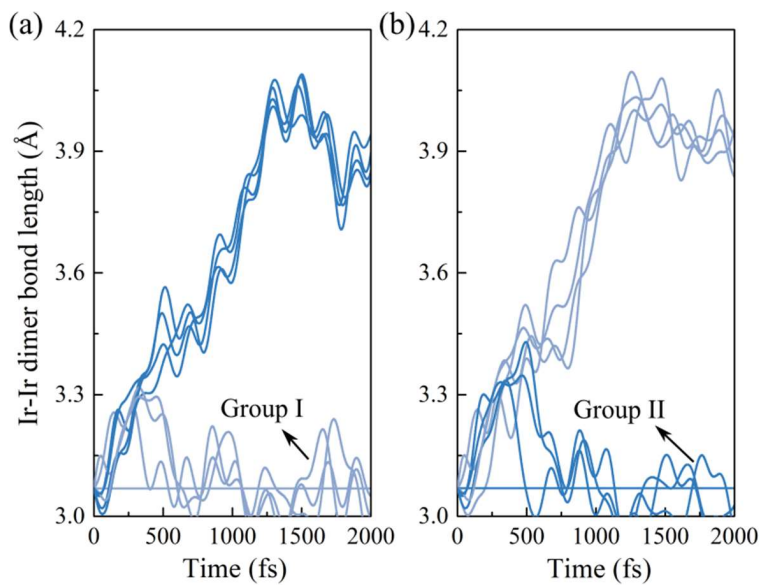

**Fig. 3. Dynamics with a fixed Ir-Ir dimer upon photoexcited 3% electrons to antibonding states.** (a) Evolution of Ir-Ir dimerized bond length with a fixed Ir-Ir dimer in group I. (b) Evolution of Ir-Ir dimerized bond length with a fixed Ir-Ir dimer in group II. The shallow (dark) blue lines represent Ir-Ir dimers in group I (II).

Surprisingly, all the dimers in group II have been dissociated, while all the dimers in group I are not dissociated, although groups I and II are equivalent in a thermodynamic average sense. Thus, the different results between groups I and II must be random results due to the thermodynamic fluctuation of the lattice. To confirm this, with the same photoexcitation, we have re-done different simulations with different random seeds for the lattice vibration. Indeed, in some other cases, the opposite happens, e.g., all the dimers in group I have dissociated, while dimer in group II has not dissociated. When one Ir-Ir dimer in group I (or group II) is fixed during the simulation, we find that the fixed Ir-Ir dimer directly prevents the dissociation of the other three Ir-Ir dimers in the same group, but the dimers in the other group are all dissociated, as shown in Fig. 3(a) and 3(b). We thus deduce that dimer dissociations are correlated in the same group and anti-correlated in the opposite group. Due to the Te atoms making an interlayer connection adjacent to the two Ir-Ir dimers within the same group (Fig. 1a), the movement of one Ir-Ir dimer will directly exert a force on adjacent Ir-Ir dimer, thus correlating their dissociations. Besides, when one Ir-Ir dimer is fixed in one group I (or II), the timescale of dimer dissociation in group II (or I) is accelerated to 1.2 ps in Fig. 3 which is fast than the dissociation in Fig. 2(e). All these means the dimer dissociations are anti-correlated between different groups, and the fixation of dimer in one group can help the dissociation of the dimer in the opposite group.

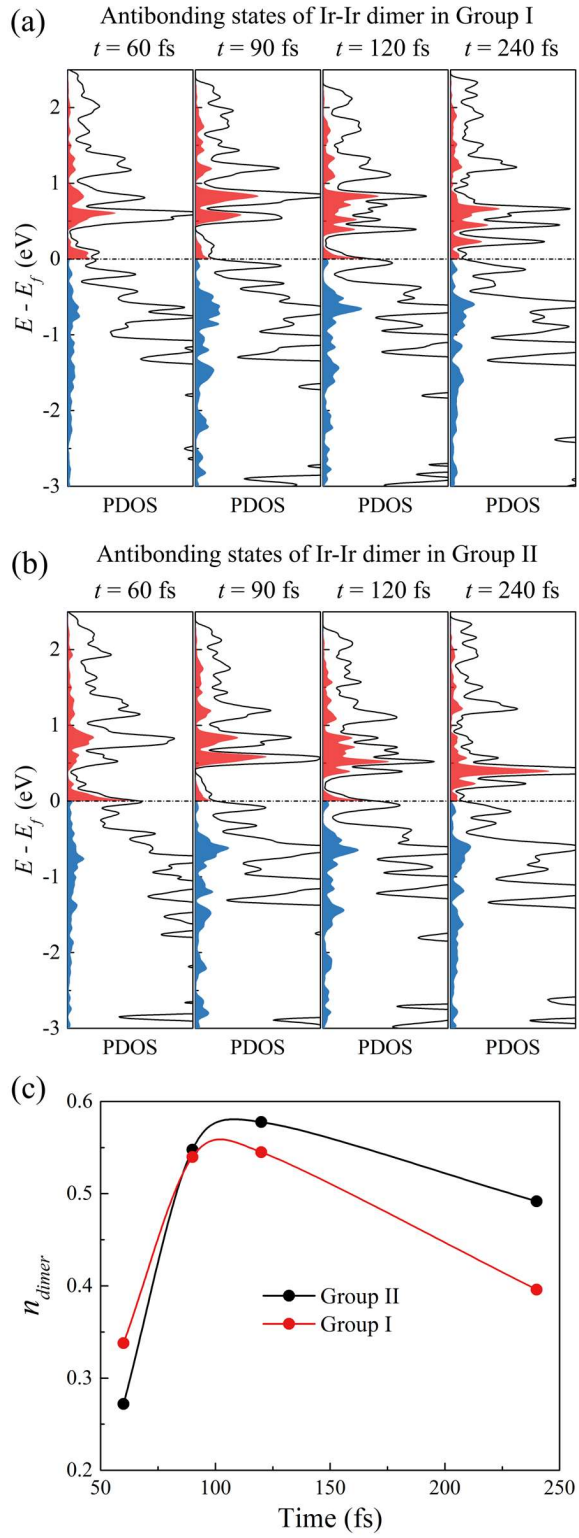

**Fig. 4. Evolution of PDOS and occupation numbers in IrTe$_2$ following photoexcitation.** (a) Evolution of Ir $d_{xz}$ + $d_{yz}$ PDOS of the Ir-Ir dimers in group I. (b) Evolution of Ir $d_{xz}$ + $d_{yz}$ PDOS of the Ir-Ir dimers in group II. Red (blue) shaded areas show the excited electrons (holes). (c) The $n_{dimer}$ is defined as the integration of the time-dependent electronic occupations of Ir-Ir dimerized antibonding states within 0.3-1.0 eV above the Femi level: $n_{dimer} = \int_{0.3}^{1.0} \rho_{dimer}(E,t)\,dE$.

Figures 4(a) and 4(b) show the evolution of the Ir $d_{xz}$ + $d_{yz}$ PDOS and the excited electronic occupation in groups I and II, respectively. To visualize the electronic occupations in Fig. 4(a) and 4(b), we proceed with the integration of the time-dependent electronic occupations of Ir-Ir dimerized antibonding states within 0.3-1.0 eV above the Femi level: $n_{dimer} = \int_{0.3}^{1.0} \rho_{dimer}(E,t)\,dE$, as shown in Fig. 4(c). The difference of the electronic occupations in groups I and II is likely introduced by the initial random fluctuation, as the laser in-plane electric field should not break their symmetry. Note, there is an instability and self-amplification mechanism which amplifies the small initial difference. This is because an initially slightly larger occupation of the anti-bonding state will elongate the dimer bond distance, which lowers the anti-bonding state energy and in-term attracts more electron transfer to this antibonding state. This thus forms a positive feedback loop. As a result, an initial small difference can lead to a random selection of either group I or group II to be dissociated.

Figure 5 summarizes the coherent and disordered dynamics of intralayer Ir-Ir dimers. The ultrafast coherent phase transition for all dimers can be achieved with the equally excited two dimer elongating modes for the two dimers in the same plane. In some sense, the two modes are against each other as one dimer bond length increase tends to suppress the other dimer bond length. In the coherent case, the two modes (or say the driving force) are the same (the thermal random force is small). Such balance results in a coherent dimer dissociation for all the dimers. The driving force is large enough to dissociate both dimers, at the cost of suppressing the middle parts (non-dimerized part) shown in Fig.5(a). In the laser excitation case, the dimer dissociation driving force is not large enough to dissociate both dimers. As a result, one of the dimers (e.g, the left one shown in Fig.5(b)) has slightly larger electron occupations for the antibonding states (due to initial random thermal vibration) will dissociate faster and collect more charge in the self-amplification process discussed above, thus at the end, only one dimer group is dissociated. This explains why the dimer dissociation in different groups is anti-correlated, and there is a large (amplified) randomness in the phase transition process. Ultrafast disordered dynamic inducing the phase transition has been reported in the experiment for VO$_2$ [18,19]. It is thought the disorder causes the different V-V to be dissociated uncorrelatedly. Here, we found that, under such a disorder process, there is a short-range order between different dimer groups. This provides new insight into the complex process of PIPTs and the de-dimerization process.

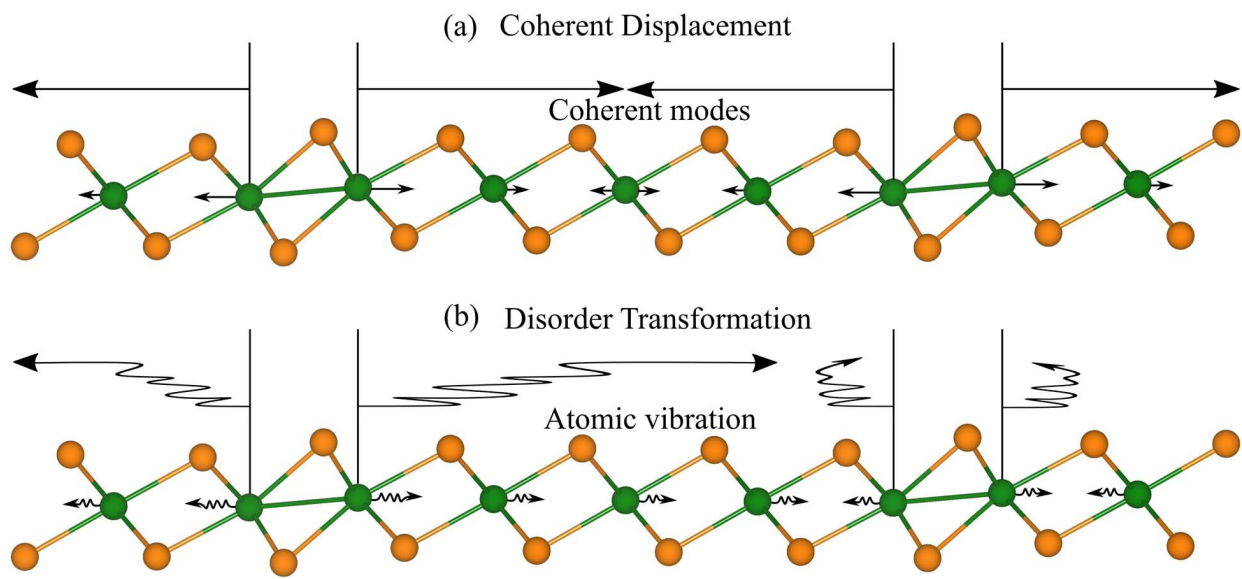

**Fig. 5. Correlated dynamics of intralayer Ir-Ir dimers in IrTe$_2$.** (a) Coherent displacement. (b) Disorder transformation.

**Conclusion**

In summary, we obtain the coherent and disordered Ir-Ir dimerized dissociation with electronic excitations in our rt-TDDFT simulations. When the Ir-Ir dimerized antibonding states are 100% occupied by the excited electrons, all the Ir-Ir dimers will be coherently dissociated in 300 fs. On the other hand, in the laser excitation case, the Ir-Ir dimerized antibonding states are only 40% occupied, and other electrons are excited to the higher-energy electronic states. Due to the smaller de-dimerization driving force, combined with the disorder thermal vibrations, the dimers will only be partially dissociated, and it will only happen at a much longer timescale of about 2.0 ps. More surprisingly, in this disordered de-dimerization, there is a local correlation between the de-dimerization between different groups (defined in Fig.1). The dimers in the same group will have a positive correlation, i.e, either they are all dissociated, or not dissociated, while the dimers in the neighboring group have anti-correlation, i.e, if the dimers in one group are all dissociated, the dimmer in the neighboring group will not be dissociated, and vice versus. We believe our results provide a new understanding of the PIPTs process.

**Acknowledgments:**

The work in China was supported by the Key Research Program of Frontier Sciences，CAS under Grant No. ZDBS-LY-JSC019, CAS Project for Young Scientists in Basic Research under Grant No. YSBR-026, the Strategic Priority Research Program of the Chinese Academy of Sciences under Grant No. XDB43020000, and the National Natural Science Foundation of China (NSFC) under Grant Nos. 11925407 and 61927901. L.W. W was supported by the Director, Office of Science, the Office of Basic Energy Sciences (BES), Materials Sciences and Engineering (MSE) Division of the U.S. Department of Energy (DOE) through the theory of material (KC2301) program under Contract No. DEAC02-05CH11231.